\documentclass[12pt,a4paper]{article}
\usepackage[toc,page]{appendix}
\usepackage{fullpage}
\usepackage[latin1]{inputenc}
\usepackage[french,english] {babel}
\usepackage{amsmath}
\usepackage{amsfonts}
\usepackage{amssymb}
\usepackage{graphicx}
\usepackage{epstopdf}
\usepackage{color}
\usepackage{listings}
\usepackage[stable]{footmisc}
\usepackage{verbatim}
\usepackage{subfig}
\usepackage[top=2.5cm,bottom=2.5cm,right=2.5cm,left=2.5cm]{geometry}
\usepackage{amsthm}
\usepackage{feynmp}
\usepackage{array}
\usepackage[normalem]{ulem}
\definecolor{vert}{rgb}{0,0.5,0}
\newcommand{\beqn}{\begin{eqnarray}}
\newcommand{\eeqn}{\end{eqnarray}}
\newcommand{\beqas}{\begin{eqnarray*}}
\newcommand{\eeqas}{\end{eqnarray*}}
\newcommand{\beq}{\begin{equation}}
\newcommand{\eeq}{\end{equation}}
\newcommand{\bit}{\begin{itemize}}
\newcommand{\eit}{\end{itemize}}
\newcommand{\bseq}{\begin{subequations}}
\newcommand{\bal}{\begin{align}}
\newcommand{\eal}{\end{align}}
\newcommand{\eseq}{\end{subequations}}
\newcommand{\nn}{\nonumber}

\begin{document}
\begin{flushright}
{ULB-TH/15-07}                      \\
INR--TH--2015-012
\end{flushright}
\vskip 1cm
\begin{center}
{\huge Flavour changing $Z'$ signals in a 6D inspired model}
\vskip .7cm
{\large Jean-Marie Fr\`ere${}^a$},
{\large Maxim Libanov${}^{b,c}$},
{\large Simon Mollet${}^a$} and
{\large Sergey Troitsky${}^b$}
\vskip .7cm
\emph{
${}^a$ Service de Physique Th\'eorique, Universit\'e Libre de Bruxelles, ULB, 1050 Brussels, Belgium
\vskip 0.2cm
${}^b$ Institute for Nuclear Research of the Russian Academy of Sciences 117312, Moscow, Russia
\vskip 0.2cm
${}^c$ Moscow Institute of Physics and Technology 141700, Dolgoprudny, Moscow Region, Russia
}
\end{center}

\begin{abstract}
We consider the phenomenology of new neutral gauge bosons with flavour
non-diagonal couplings to fermions, inherent in 6D models
explaining successfully the hierarchy of masses as well as the mixing for quarks, charged leptons
and neutrinos (this model can in particular be credited with the  correct prediction of the neutrino mixing angle $\theta_{13}$). We present a general relation between masses of new gauge
bosons and their couplings to fermions.
We show that in the current realization of the model, the new heavy bosons are unreachable at LHC but argue why the constraint could be relaxed in the context of a different realization. In  view of a more systematic study, we use an effective model inspired by the above to relate directly rare meson decays to possible LHC observations. In terms of effective Lagrangians, this can  be seen as the introduction in the model of only one overall scaling parameter to extend our approach without modifying the 4D (gauge) structure.
\end{abstract}

\section{Introduction}

Models that reach beyond the Standard Model (SM) usually introduce new
particles, among them neutral gauge bosons. Canonically, neutral gauge
bosons tend to be flavour-diagonal through a generalization of the
Glashow-Iliopoulos-Ma\"{\i}ani mechanism. An exception arises when family
symmetries are involved. Groups involving "horizontal symmetries" are an
obvious example, but we have shown~\cite{FLNT_flavour} that a similar
(although usually less dramatic) situation obtains in models with extra
dimensions (ED), when the family replication itself is associated with the
extra spatial dimensions~\cite{LT, FLT}. In particular, the
lowest mode of the "Kaluza-Klein" tower remains flavour-diagonal (at least
to an excellent approximation), but higher excitations can show either
a departure from the Cabibbo-Kobayashi-Maskawa unitarity, or an outright
violation of flavour conservation in neutral currents. In a
six-dimensional (6D) model we have studied, some of the excitations
effectively carry "family number" and mediate flavour changing (but
approximately family-conserving) neutral transitions. These higher modes
could become detectable in precision low energy
processes~\cite{FLNT_flavour, Libanov:2012kh, Shkerin:2015mea} or even at
colliders like the LHC~\cite{FLNT_LHC}.

Of course, different models give various spectra of particles with their
own features each. Here, we would like to come back on a model we
developed during the last decade and which proved fruitful to explain
masses and mixing of both charged and neutral
fermions~\cite{Frere:2013eva, LN_fit, FLL_neutrino, Frere:2014rra} (in particular, we
made predictions for the neutrino
sector~\cite{FLL_neutrino,Frere:2014rra, Lnu-talk}, relating Majorana character to large
mixing and inverted hierarchy, and successfully predicting -- \emph{in
tempore non suspecto} -- the last mixing angle $\theta_{13}$). In
this model, we add to our usual four-dimensional (4D) world, two ED with a
Nielsen-Olesen vortex structure on it. Quite generally in this background,
we can get $n$ localized (chiral) fermionic zero modes from a single
spinor in 6D \cite{LT, FLT, JackiwRossi, FLNT_sphere}; these zero modes
(choosing the vortex winding number $n=3$) then play the role of the different generations of SM
quarks and leptons.

However, the requirement of both a normalizable zero mode for gauge bosons
and of charge universality tends to restrict the arbitrariness of geometry
in ED. In our recent review (\cite{Ru60} and references therein), two
archetypal approaches were studied: either we can work in a compact space
(with the sphere as the typical prototype) or in a warped space
\textit{\`a la Randall-Sundrum}\footnote{In this case, the prototype is a
warped plane whose geometry is stabilized by the vortex structure.}.
There, we also argue that both solutions should give a reasonably similar
phenomenology and so we will stick here with the first one that we have already carefully analysed (see~\cite{Frere:2013eva} for a complete review).
Up to now, this analysis was restricted to its simplest form --- \textit{i.e.} the spherical compactification --- which unfortunately, as we will show, offers a poor framework for LHC phenomenology. It would be desirable to go  beyond and explore more complex geometries.
For the time being, we take an effective approach, where these possible extensions are parametrized in term of a single new overall scaling factor (in which we suppose are embedded the details of the geometry) or, and this is closer to our current approach, are directly studied in the context of 4D low-energy phenomenological models patterned after the 6D original one.

In section \ref{sec:2}, we will remind the basics of our model in the
gauge sector. Section \ref{sec:3} deals in some more detail with mixing
effects. We discuss in section \ref{sec:4} the phenomenology (at LHC) and conclude in section
\ref{sec:5}.

The present work differs in significant points from the work
of~\cite{FLNT_LHC}. Notably, the effects of flavour mixing
were not included in the previous work. Instead of considering $Z'$ and
$\gamma'$ modes, we argue here that the $W_3'$ and $B$ modes, summed
incoherently, should be used.

\section{Gauge bosons sector}\label{sec:2}

In the kind of models we deal with, each mode (field excitation) is
associated with a wave function in the ED. For normalizable modes, we can
always integrate over the ED variables, what leaves us with an effective
4D theory (compactification procedure) where the different modes interact
among themselves. Separation of variables in ED allows to decompose wave
functions as a product of a radial part and an angular one\footnote{On the
sphere the colatitude $\theta$ plays the role of the radial variable.}.
The angular part is a typical Fourier expansion $e^{i\nu \varphi}$ where
$\nu$ is called the "winding" number. Then, the integration over the
radial component controls the strength of the interaction through the
overlaps of wavefunctions (see below), while the one over angular
component gives an obvious selection rule which forbids interaction with
nonzero total winding\footnote{This can be interpreted as the angular
momentum conservation in the ED; these selection rules are however
violated when dealing with family mixing.}.

Our low-energy fermions have localized radial wave functions and winding
numbers of successive generations differing by one unit. On the other
hand, gauge-boson wave functions are usual spherical harmonics
propagating all over the ED. They are characterized by the couples $(\ell,
m; 0 \leq \ell,\vert m\vert \leq \ell)$. For a given $\ell$ there are $(2\ell
+1)$ 4D modes with mass $\sqrt{\ell(\ell+1)}/R$ ($R$ is the sphere
radius). Thus, the mode $\ell=0$ plays the role of the usual SM boson.

Now let us quickly come back on the resulting 4D interactions between SM
fermions and (neutral) bosons modes. As established in previous papers\footnote{For the details of the calculations of the effective SM
Lagrangian see, \textit{e.g.},\cite{FLNT_flavour},  properties of the
Brout-Englert-Higgs boson are discussed in~\cite{Frere:2013eva,Libanov:2007zz}.}, for any neutral (6D) gauge field $W_A$ which interacts with the fermions,
we get the following effective Lagrangian at 4D level\footnote{Note that
family mixing through the matrices $U$ breaks the conservation of winding
number, as a fermion mass eigenstate is now a mixing of wave functions
with different winding.}:
\beq \mathcal{L}
_{4D} \supset \sum_{\ell}\sum_{\underset{\vert n-m\vert \leq \ell}{m,n}}
E^{\ell,\vert n-m\vert}_{mn}U^*_{mj}U_{nk}\left(\bar{\psi}_j \gamma^{\mu}
Q \psi_k\right) \omega^{(*)}_{\mu;\ell, \vert n-m\vert}
\label{Eff4D}
\eeq where $E^{\ell,\vert n-m\vert}
_{mn}$ are the results of the overlaps (see below). For $\ell=0$, we have
$E^{0,0}_{nn}=1$ (normalization) which permits to identify $Q$ with SM
charges. $U$ is the unitary mixing matrix%
. While it disappears properly for $\ell=0$ (thanks to perfect
unitarity), this is no more the case for higher $\ell$'s. Thus, in our
model, it makes sense to talk about mixing in up quarks and down quarks
separately, for instance. $\omega^{(*)}_{\mu}$ are the 4D fields for each
mode ($Z'$-like bosons for $\ell\neq 0$). When $n-m\neq 0$ these are
complex fields. In our notations, for $n-m>0$ we have to use
$\omega_{\mu}$, so it destroys a mode with winding $\vert n-m\vert$, while
for $n-m<0$ we have to use $\omega^*_{\mu}$, so it creates a mode with
winding $\vert m-n\vert$.

We remember the big point of our model: even without mixing ($U=1$) we
have $Z'$-like bosons\footnote{We use generically $Z'$ for any uncoloured neutral gauge boson; it could stand for $W'_3,B',...$} that mediate flavour changing processes.
Nevertheless, in this first approximation (no mixing) only processes with
$\Delta G=0$ (where $G$ is some kind of family number) are allowed. In
fact, in this approximation, the transition $d +\overline{s} \rightarrow
Z' \rightarrow  e^- + \mu^+$ is flavour-changing (and family number
conserving), but the corresponding $d +\overline{s}\rightarrow Z'
\rightarrow d +\overline{s}$ is purely flavour-conserving (and does not
contribute thus to $K^0 - \overline{K^0}$ mixing). We have already
extensively commented on the role of the mixing in~\cite{FLNT_flavour}.
Here we will rather come back on the overlaps and their link to the
geometry but below we will give an example of the treatment of mixing in a
concrete application.

In~\cite{FLNT_flavour} overlaps $E^{\ell,\vert n-m\vert}_{mn}$ have been
estimated for fermions sufficiently localized ($\theta_f \lesssim 1$,
where $R\theta_f$ is a measure of the fermionic extension in ED). The
results were:
\beqn E^{\ell,\vert n-m\vert}
_{mn} \sim \left\{
\begin{aligned}
& \sqrt{\ell} (\ell\theta_f)^{\vert m-n\vert} \quad \text{at }
\ell\theta_f \ll 1,\\
& \frac{1}{\sqrt{\theta_f}}\quad\quad\quad\quad\ \ \text{at }
\ell\theta_f \simeq 1.\\
\end{aligned}
\right.
\label{Overlaps}
\eeqn Note that for $\ell\theta_f\gg 1$ (higher modes), the fast
oscillations in the region of significant overlap with the fermions
quickly cut-off the integral. In the following we will neglect these
modes. We emphasize the continuity between the regimes $\ell \theta_f
\ll 1$ and $\ell\theta_f \simeq 1$. It is also worth noting that the
strongest couplings between fermions and massive gauge modes are for those
with $\ell\sim 1/\theta_f$ (what is obvious in terms of overlap).
This is a quite general feature, already mentionned
in~\cite{FLNT_flavour}, but on which we have not insisted in~\cite{Ru60}.
For this reason, rare decays (like $K_L^0$ decay) tend to bound directly
the Kaluza-Klein scale (the mass of the first excitation) of the bosons
(the size $R$ of the sphere) rather than some combination of the mass and
the overlap. Of course, the higher modes are suppressed by large $Z'$ masses, but this is insufficient to compensate for the large overlaps and number of possible exchanges. Indeed, in $K_L^0$ decay, if we allow for all modes to be
exchanged (at least until the natural cut-off provided by fast
oscillations) in the process, this gives for the decay amplitude (up to standard numerical
factors) $\mathcal{M}\sim R^2\sim 1/M^2_{Z'}$ (where we have defined
$M_{Z'}$ as the mass of the first massive excitation
$\ell=1$)\footnote{Note that the same order of magnitude is obtained
either if we sum over all modes starting with $\ell=1$ to
$\ell=\ell_{\text{max}}\sim 1/\theta_f$ or if we consider all the
contributions of the same size as the dominant one
($\ell=\ell_{\text{max}}$) -- the saddle-point approximation.}.

With the prospect of looking into more phenomenological models (see
below), it is still interesting to consider the couplings of the lower
modes, where we observe an interesting fact:
\beq \frac{E^{\ell,\vert n-m\vert}
_{mn}}{M_{Z'}^{\vert n-m\vert}}\sim l^{\vert n-m\vert
+1/2}(R\theta_f)^{\vert n-m\vert}\sim C_{\ell,\vert n-m\vert,f}\label{overlapSc}
\eeq The last result means that these ratios are some constant depending
only on the mode and not on physical mass (which is a function of $R$).
This is due to the fact that the physical size of the fermion
$(R\theta_f)$ is fixed once and for all by the coupling to the vortex that
has nothing to do with the present discussion. In particular
$E^{\ell,1}/M_{Z'}=$ const and $E^{\ell,0}=$ const.

This is phenomenologically very interesting, because we get a relation
between the mass of new $Z'$ bosons and their couplings with fermions. In
particular, a flavour changing $Z'$ current will couple with a strength
proportional to its mass.

To summarize this section, we have considered here in addition to the
"canonical" case where the vortex and the fermions take a large fraction
of the sphere, the situation where our fermions are concentrated on a very
central region of the vortex. Imagine for instance that the region
occupied by the 4D fermions is, say, $1/100$th of the sphere; we have seen
that in this case, the coupling of the lowest-lying flavour-changing gauge
bosons is suppressed by a corresponding factor $\simeq 1/100$ (see
Eq.~(\ref{Overlaps}) with $\ell=1$).  However, in the same situation, the
mode $\ell=100$ would have full unsuppressed (and large, $E\sim 10$; see
Eq.~(\ref{Overlaps})) coupling to the fermions, and corresponding
transitions would only be suppressed by its mass. In practice, this shows
that, if the full Kaluza-Klein tower is allowed to contribute, the
fraction of the sphere occupied by the fermions does not impact the limits
obtained from rare ($K$ or other meson) decays on the Kaluza-Klein mass
$1/R$.

Typically, this leads to a $Z'$ of mass $\cong 71$ TeV, out of reach of current or planned colliders.
We will however pursue in  section \ref{sec:3}  the discussion of flavour-changing $Z'$ contributions, both for its theoretical interest, and in preparation for section \ref{sec:4} where we will relax some of the constraints, leading to accesible $Z'$  and study possible LHC phenomenology.

\section{Mixing effects}\label{sec:3}

As we see from the previous section, we have a model which predicts
$Z'$-like bosons, some of which mediate peculiar flavour changing
processes. Moreover, we predict the relation between their masses and
couplings to fermions. If we take the model at face value (accepting the
high cut-off mentioned before), constraints from meson decays make it
untestable by current (and probably even future) colliders. We reach indeed
a lower bound of $1/R \gtrsim 50$~TeV, translated into 
\beq
M_{Z'} \gtrsim 71 \text{TeV}\label{bound}
\eeq
This bound was already mentioned in \cite{Frere:2013eva} where fermions are sufficiently wide such that the dominant contribution in $K_L$ decay (Br$(K_L\rightarrow \mu^+e^-)\leq 4.7 \cdot 10^{-12}$ \cite{PDG}) comes from the lower modes (see \cite{Ru60}).
We have now seen in section \ref{sec:2} that modifying the localization of the fermions (for instance, tightly around the origin) does not modify
this limit: while the lower modes contributions are indeed suppressed, significant contributions to the rare $K$ decays are instead dominated by the high $\ell$ modes.

Despite this, we pursue with the estimation of the respective "diagonal" and "off-diagonal" $Z'$ contributions to flavour-changing processes
in a collider context, as these will remain valid in the extended models considered in section \ref{sec:4}.

What we are mainly interested in are production of $m=1$ bosons whose
typical signature would be a lepton-antilepton pair ($e\mu$) or
($\mu\tau$) with large and opposite transverse momenta. This is very
similar to Drell-Yan pair production for which a typical feature is the
suppression of the cross section with increasing of the resonance mass at
a fixed center-of-mass energy. Note also that, since we are dealing here
with proton-proton collisions, we expect a dominance of $(e^-\mu^+)$ and
$(\mu^- \tau^+$) over $(e^+\mu^-)$ and $(\mu^+\tau^-)$. Indeed the former
processes can use valence quarks ($u$ and $d$) in the proton, while the
latter involve only partons from the sea.

However, we must be careful for it could be that, because of mixing, a
"diagonal" $Z'$ (\textit{i.e.} a mode with $m=0$) would produce the same
signature that a true "flavoured" $Z'$ ($m=1$ mode) or even dominate it
(actually we have seen that while couplings with $m=1$ modes scale with the
mass, couplings with $m=0$ modes stay approximately constant so the
question is nontrivial).

Let us analyse the $(e^-\mu^+)$ production. When the mixing is taken into account, the
interactions with $\ell=1$ bosons are (see Eq.~(\ref{Eff4D})):
\beqn \mathcal{L}
&\supset &
\left\lbrace(E_{12}^{1,1}U^*_{11}U_{22}+E_{23}^{1,1}U^*_{21}U_{32})\
\omega_{\alpha;1,1}\right.\nn\\
&+&
(E_{11}^{1,0}U^*_{11}U_{12}+E_{22}^{1,0}U^*_{21}U_{22}+E_{33}^{1,0}U^*_{31}U_{32})\ \omega_{\alpha;1,0}\nn\\
&+& \left.(E_{21}^{1,1}U^*_{21}U_{12}+E_{32}^{1,1}U^*_{31}U_{22})\
\omega_{\alpha;1,-1}\right\rbrace
(\bar{e}\gamma^{\alpha}Q\mu)
\label{effLag1}
\eeqn In absence of mixing, only the interaction with $m=1$ mode survives.
Others are thus suppressed by the mixing. As an important example, we have compared $pp\rightarrow \mu^+e^-$
cross-sections for of a $m=1$ and for a $m=0$ boson exchanged in
(thought-experiment) machines with center-of-mass energy
$\sqrt{s}>M_{Z'}=71$ TeV. Indeed, a $m=0$ mode can be produced by $(u\bar{u})$ or $(d\bar{d})$ in the protons which are more abundant than $(u\bar{c})$ and $(d\bar{s})$ and constitutes then the only other contribution that could be relevant for this issue. We have obtained (details about the calculations can be found in
Section \ref{sec:4}):
\beq
\frac{\sigma_{pp\rightarrow \omega_0\rightarrow \mu^+e^-}}{\sigma_{pp\rightarrow \omega_1\rightarrow \mu^+e^-}}
\sim 10^{-3} - 10^{-2}\label{XsecRatio}
\eeq

Nevertheless, interaction with $m=0$
mode could potentially be dangerous yet when we scale $M_{Z'}$ below $71$ TeV since we have just seen that $E^{1,0}$ factors should stay
approximately constant. Fortunately an additional
suppression is hidden in the coupling. Indeed, $E^{1,0}_{ii}$ factors
result from the overlap of fermions and gauge wave functions, the latter
being $\sim Y_{10}\sim P^0_1 \underset{\theta\sim 0}{\sim}\cos\theta$.
This means that for sufficiently narrow fermionic profiles, it can be
replaced by a constant in the integral. Therefore $E^{1,0}_{ii}\sim$
"fermion normalization" and then $E^{1,0}_{11}\simeq E^{1,0}_{22}\simeq
E^{1,0}_{33}$. In this limit, the total coupling to $m=0$ boson can be
rewritten $\simeq E^{1,0}_{11}(U^{\dagger}U)_{12}\simeq 0$, the last
equality resulting from unitarity of $U$. So "quasi-unitarity" tends to
suppress the coupling to $m=0$ boson and as we will see, it compensates
for the invariance of individual $E^{1,0}$ factors when theory is scaled to
lower $M_{Z'}$ masses.

\section{Phenomenology at LHC}\label{sec:4}

As seen before, simply trading the localization of the fermions on the sphere for a tighter one does not help (in our particular case) in lowering the limit
on the Kaluza-Klein mass.
We now depart from the canonical model (but will keep intact the structure of the 4D effective Lagrangian).

We have considered two ways in which we can envisage such departure.

\begin{itemize}
  \item We can consider the 4D Lagrangian in a strict "cut-off" limit, placing an arbitrary cut-off (this is needed anyway since the 6D theory is not renormalizable). If we choose the cut-off to be just above the first Kaluza-Klein excitation $\ell=1$, and in the same time keep the fermions thightly localized around the origin, the constraints from rare $K$ decays are considerably lowered to $M_{Z'} \geq (\hat{E}^{1,1}/\hat{E}^{1,1}_{\text{old}})\cdot 71 \text{TeV}$ where "old" designates the ancient value used in the case of wide fermionic profiles (see \cite{Frere:2013eva}) and the hat "$\ \hat{}\ $" is to keep in mind that this actually is a combination of such "flavoured" overlap factors that appears because the individual ones generally are different for distinct fermions species and/or chiralities. 
  
Nevertheless, we remind from (\ref{Overlaps}) that $E^{1,1}\sim \theta_f$, the portion of the sphere occupied by the fermions. We can now indeed lower the bound (\ref{bound}) by localizing the fermions on a tighter region and thus reach any $M_{Z'}$ below 71 TeV. Remember that (see equations (\ref{Overlaps}) and (\ref{overlapSc})), keeping everything else unchanged, this will corresponds to a scaling $E^{1,1}_{\text{old}}\rightarrow E^{1,1}=\kappa E^{1,1}_{\text{old}}$ with $\kappa \propto M_{Z'}$. Then, to saturate the bound, it suffices to choose:
  \beq
  \kappa=\frac{M_{Z'}}{71 \text{TeV}}\label{kappa}
  \eeq   
  \item Alternatively, we can keep in mind that we have only explored a limited set of geometries in 6-D, and consider that more general cases could lead to different value of the overlaps between $Z'$ and fermions. While keeping the model intact, we introduce a parameter $\kappa$ to explore this overlap suppression. This will of course lower the bound in the same way: $M_{Z'}\geq \kappa \cdot 71 \text{TeV}$ and it suffices again to choose $\kappa$ as (\ref{kappa}) to saturate it.
\end{itemize}

In both approaches, we keep a highly constrained Lagrangian, inherited from the mass generation mechanism, and introduce an extra parameter to take into account either or ignorance of 6D structure, or simply to account for the fact that the 4D theory is only an effective one, requiring an explicit cut-off\footnote{For the sake of consistency, we will also try to take the "quasi-unitarity" suppression discussed in section \ref{sec:3} into account.}.

As
such, this effective model (or, in current parlance a "simplified model")
allows to compare the constraints from precision measurements (Kaon
decays) to the reach of LHC for flavour-changing but (nearly)
family-number conserving vector bosons. Note that in both cases, Kaon limits impose that the coupling of the $Z'$ to the fermions is scaled precisely by the factor $\kappa$, thus trading lower $Z'$ masses for suppressed production cross sections, and narrower width.

In this context, for sufficiently small $\kappa$, $Z'$ bosons become accessible at LHC, but we must still consider if they will be produced
in sufficient numbers.
We also face some interesting questions concerning the coherence (or lack thereof) of the various contributions, and the nature ($W'_3 - B'$ or $Z' - \gamma'$)
 of the Kaluza-Klein modes.

 At energies
$\gtrsim M_Z$, it seems reasonable to use $W'_3$ and $B'$ rather
than $Z'$ and $\gamma'$, as corrections to the masses due to gauge
interaction are expected to overwhelm those due to electroweak symmetry
breaking. For all these $\ell=1$ bosons, the ratio width/mass at $71$ TeV
is of the order $10^{-3}-10^{-2}$. Indeed, from Eq.~(\ref{effLag1}), neglecting the
fermions masses, we get:
\beqn
\Gamma_{W'_3}&=&\frac{g^2M_{W'_3}}{48\pi}\left(A_L+N_CA_Q\right),\nn\\
\Gamma_{B'}&=&\frac{g'^2M_{B'}}{24\pi}\left(2y_l^2A_L+y_e^2A_e+N_C(2y_Q^2A_Q+y_u^2A_u+y_d^2A_d)\right),\nn
\eeqn where $N_C=3$ is the number of colors, $y$ are the weak
hypercharges\footnote{We use the definition $Q=T_3+Y$.} and
$A=\sum_{i=1}^3 (E_{ii}^{1,0})^2$ for $m=0$ bosons or
$A=(E_{12}^{1,1})^2+(E_{23}^{1,1})^2$ for $m=1$ bosons. For $m=0$ bosons we
get $\Gamma_{W'_3}=2.67$~TeV and $\Gamma_{B'}=1.33$~TeV at $71$ TeV; while
for $m=1$ bosons we have $\Gamma_{W'_3}=0.75$~TeV and $\Gamma_{B'}=0.48$~TeV
\footnote{This corrects the smaller values erroneously mentioned in
\cite{Ru60}.}. Note that at lower masses, the ratio $\Gamma/M$ will even be reduced further (for $m=1$ bosons, see below).
It means that even with a small separation of both $W'_3$ and $B'$ masses
(due to running and/or EW symmetry breaking), we could neglect
interferences between the two channels to a good approximation. We have
thus in the following calculation used $W'_3$ and $B'$ and added the
contributions incoherently. In order to minimize the number of parameters
we have however not taken into account the mass difference of $W'_3$ and
$B'$ in the final result (while still insisting on the incoherent
sum)\footnote{On the other hand, we have take the running of the EW
coupling constants into account for the widths and the cross-sections
calculations.}. To simplify the calculation, we also use the narrow-width
approximation for the propagators.

The individual cross-sections will always be of the form:
\beq \sigma\sim \frac{C}{M\Gamma},\nn
\eeq where $M$ and $\Gamma$ are the boson mass and width respectively, $C$
is a numerical factor that contains the coupling at the fourth power and depends only on $M$ through parton distribution functions (pdf)\footnote{We use MSTW parton distribution functions \cite{Martin:09} at leading order (we have checked that corrections next to the leading order have negligible effect).}.

In Section \ref{sec:3}, we have computed this for $m=1$ and $m=0$ bosons
and showed that the first one clearly dominates. Now, we know that the
coupling for $m=1$ will scale with the mass while the $m=0$ coupling will
only scale thanks to "quasi-unitarity" at the letpon vertex. Let us define
parameters $\kappa=M/(71~{\rm TeV})$ and $\delta$ which encodes the
unitarity reduction.

\bit
\item For $m=1$, $\Gamma\sim \kappa^2 M= M^3/(71\text{ TeV})^2$ and
$C\sim \kappa^4=M^{4}/(71\text{ TeV})^4$, so we have:
\beq
\sigma\sim \frac{1}{(71\text{ TeV})^2}\nn
\eeq
The only change comes from the particle distribution functions (pdf).
\item For $m=0$, $\Gamma\sim M$ and $C\sim \delta^2$, so we have:
\beq
\sigma\sim \left(\frac{\delta}{\kappa}\right)^2\frac{1}{(71\text{ TeV})^2}\nn
\eeq
Here because of the mass reduction (encoded in $\kappa$ for convenience),
the cross-section has the tendency to increase, but this is largely
compensated by $\delta$. As an example, we have computed $\delta\simeq
2\cdot 10^{-3}$ at $1.5$ TeV for a $W'_3$ exchange. For this mass,
$\kappa\simeq 2\cdot 10^{-2}$. So we have indeed a suppression of the
order $10^{-2}$ in addition to the original suppression at 71~TeV (\ref{XsecRatio}). We
can \textit{a fortiori} neglect this contribution in a very good
approximation at lower masses. \eit

\begin{figure}[h]
\centering
\resizebox{1.0\textwidth}{!}{\input{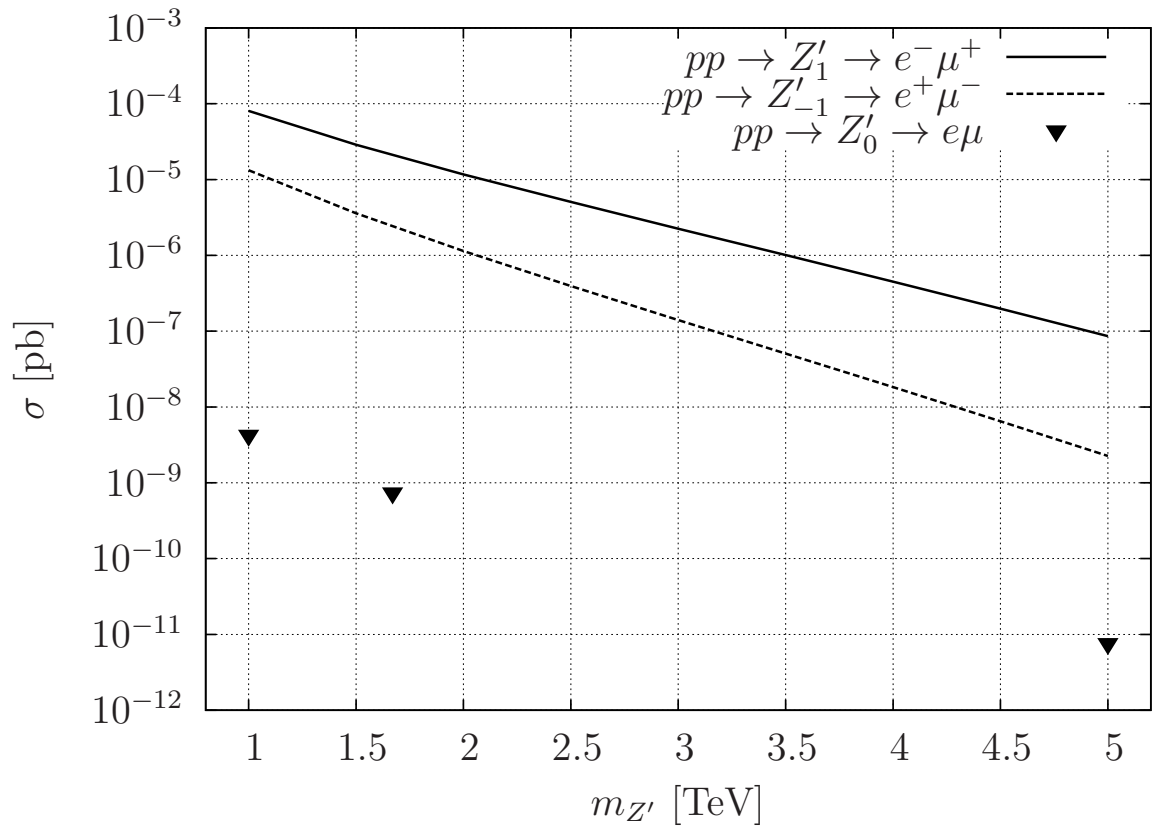}}
\caption{\label{fig:LHCpred} Predicted cross sections (at LO) of $\mu^{+}e^{-}$
(full line) and $\mu^{-}e^{+}$ (dashed line) production at the LHC
(13~TeV) versus the $Z'$ boson mass; the lines represent the upper bounds imposed by $K_L$ decay. The triangles give an estimation of the same production through a $\ell=1,m=0$ boson which is indeed suppressed by $4$ or $5$ orders of magnitude.}
\end{figure}

Figure \ref{fig:LHCpred} shows our predictions for production
cross-sections for $\mu^+ e^-$ and $\mu^- e^+$ at $13$ TeV. The lines represent the upper bounds imposed by $K_L$ decay. As expected,
the first one is dominant.

\subsubsection*{Is this the right Z' boson~?}

Assuming evidence for such a $Z'$ boson, or rather for such a family of $Z'$ bosons (carrying family number $-1$, $0$, and $+1$, like the 6$D$ model predicts it), how can we be sure this is indeed the manifestation of our $6$D model and not a more banal heavy vector boson introduced in a renormalizable $4$D model~?

The answer is manifold. One of the simplest ways to distinguish the 2 situations is of course in the case where we put our cut-off above the $n$-th recurrence, with $n>1$; observing a similar family of neutral bosons, this time with family number ($-2$,$-1$,$0$,$+1$,$+2$) at twice the mass would of course be convincing, but this assumes that such an energy can be reached, which motivates us to look for further discriminations.

The most important point is the gauge structure. For simplicity, let us discuss first the "$B$" mode ($B$ refers to the gauge boson associated to U(1) in the SM, or rather to the associated Kaluza-Klein tower). If we limit ourselves to the fundamental and the first excitation, we get ($B, B'_{-1}, B'_{0},B'_{+1}$), where the indices refer to the amount of family number carried by the particle. Since we are dealing with a U(1) structure, those massive bosons (their mass is mainly due to the Kaluza-Klein tower) don't have mutual interactions.

This is to be contrasted with a "horizontal" symmetry between families (notably in $4$D). In such a symmetry, we can think of the families being a triplet of some "horizontal" SU(2)$_X$, with an associated gauge triplet ($X_{-1},X_{0},X_{+1}$) carrying family number. In this case, the mass of the gauge triplet would come from some additional Brout-Englert-Higgs mechanism (for instance an SU(2)$_X$ doublet scalar, which would behave as a singlet under the SM groups). The scalar structure could then be seen, but more importantly, the gauge bosons $X$ would interact among themselves according to the SU(2)$_X$ structure coefficients.

Of course, such an horizontal group has been considered many times in the past, notably in the context of "Extended Technicolor", with a larger group, and in the hope to feed the mass of the observed quarks from that of hypothetical "Techni-quarks". This attempt mostly failed, for gauge bosons light enough to feed down the masses would have implied unacceptable flavour changing neutral currents (FCNC). There is certainly no objection to having an horizontal group with a sufficiently massive scale (if we introduce it in an ad-hoc way), if we don't task it with providing the fermion masses. Still in that case, as mentioned above, its physical properties (and in particular self-interactions) differ completely from the structure arising from our $6$D model, where family number is simply associated to rotation in the extra two dimensions.

The argument made here for the $B'$ holds of course for the other Kaluza-Klein excitations. Namely, the ($W'_{i;-1},
W'_{i;0},W'_{i;+1}$), which are the three $\ell=1$ excitations of the ($W_i$) bosons, will have self interactions under the usual SU(2)$_L$, which means the index $i$, but NOT along their Kaluza-Klein numbers ($\ell=1, m=-1,0,+1$).

\section{Conclusions}\label{sec:5}

In this note, we considered in some detail the possibility that the new
physics behind the Standard Model may give rise to new neutral gauge
bosons whose coupling to fermions is not flavour-diagonal. This is the
case in a previously developed class of models with large extra dimensions
which successfully explains the hierarchies of masses and mixings of both
charged and neutral fermions of the Standard Model. We studied collider
phenomenology of these new bosons in various realizations of the model. We
pointed out that there exists a nontrivial relation between the mass of
the boson and its coupling to fermions, which makes our predictions quite
independent from a particular realization of the model. In particular, the
scale of the Kaluza-Klein modes, of which our $Z'$ is a representative,
is not changed drastically depending on the localization scale of fermions.

In the simplest, spherical geometry, explored in detail, the limits from
rare processes push the $Z'$ mass as high as $M_{Z'} \gtrsim 71$~TeV. Even
at the 100-TeV collider with the luminosity of $\sim 10$~ab$^{-1}$, see
Ref.~\cite{100TeVlumi}, there is virtually no chance to discover the boson
in this particular realization of the model. However, as discussed before, Nature might choose a different, yet unexplored way of
compactification where the flavour-changing $Z'$ may be within the reach
of the LHC.
This was considered, either of the result of a different geometry (suppressing the overlaps), or in an effective approach, where an explicit
cut-off limits the effective Lagrangian to the lower Kaluza-Klein modes.

 Taking the mixing into account, we obtained predictions for
the cross sections of $\mu^{+}e^{-}$ and $\mu^{-}e^{+}$ production at the
13-TeV LHC in this latter scenario. As expected, a clear signature of our
model is the dominance of $\mu^{+}e^{-}$ by one order of magnitude. This
version of the model may thus be easily tested at the LHC Run~2.

We are of course well aware that the numerical treatment presented here is very standard and lacks of accuracy for the purpose of a serious LHC study. Hence we plan to provide more elaborated results in the future. However, we think that this specific model deserves to be discussed at this level already for its original and generic features that should not depend on the precision of the tools artillery involved.

As a last comment, let us add that the model predicts also Kaluza-Klein excitations for gluons which would yield to considerably more events. Details of hadronization are however not very well known, so we postpone any analysis of the jet production.

\section*{Acknowledgements}

This work is funded in part by IISN and by Belgian Science Policy (IAP
"Fundamental Interactions").
The work of M.L.\ and S.T., related to elaboration of the
model of the origin and hierarchy of
masses and mixings in the context of new
experimental data, is supported by the Russian Science Foundation, grant
14-22-00161.
We thank Thomas Reis for discussions.

\end{document}